\documentclass[sts]{imsart}

\RequirePackage{amsthm,amsmath,amsfonts,amssymb}
\RequirePackage[numbers,sort&compress]{natbib}
\RequirePackage[colorlinks,citecolor=blue,urlcolor=blue]{hyperref}
\RequirePackage{graphicx}

\startlocaldefs
\theoremstyle{plain}

\theoremstyle{remark}

\endlocaldefs

\begin{document}

\begin{frontmatter}
\title{Detecting Outliers in Multiple Sampling Results Without Thresholds}
\runtitle{Detecting Outliers Without Thresholds}

\begin{aug}

\author[A]{\fnms{Yu-Fu}~\snm{Shen} \ead[label=e1]{shenyf@cho.ac.cn}\orcid{0000-0003-4445-6504}}

\address[A]{Changchun Observatory,
National Astronomical Observatories, Chinese Academy of Sciences \printead[presep={ ,\ }]{e1}}
\end{aug}

\begin{abstract}
Bayesian statistics emphasizes the importance of prior distributions, yet finding an appropriate one is practically challenging. When multiple sample results are taken regarding the frequency of the same event, these samples may be influenced by different selection effects. In the absence of suitable prior distributions to correct for these selection effects, it is necessary to exclude outlier sample results to avoid compromising the final result. However, defining outliers based on different thresholds may change the result, which makes the result less persuasive. This work proposes a definition of outliers without the need to set thresholds.
\end{abstract}

\end{frontmatter}
\section{Introduction}

People often determine the probability of occurrence of event through random sampling, but results are unreliable if the number of the sample is too small. The probability density function, which depends on both the number of samples and the number of events, is superior to a single probability value. Bayesian statistics goes further by emphasizing the importance of the prior distribution; for example, if there is a strong selection effect during sampling, no matter how large the sample is, the result still be unreliable. However, it is difficult to obtain an appropriate prior distribution in practical situations, and sometimes we may not even be aware of the selection effect during sampling, mistakenly assuming that all samples are equally weighted. This is especially common in social investigations and astronomical spectroscopic surveys. All investigations or observations have different environments, it is difficult to assess the selection effect of each of them. Bayesian linear statistics\cite{2008Bayes} take this issue into account, but this work argues that some sampling results with strong selection effects should be identified first, and they can be defined as outliers.

Sometimes sampling results with strong selection effects may be identified manually. Regardless of whether they can be manually identified, if one wants to exclude some sampling results, they must either find clear evidence of the problem within these sampling results or classify them as outliers using a strict definition, otherwise there may be suspicion of cheating. Of course, sometimes the majority of samples make the same error, and the best sampling results may end up being outliers.

The method based on the standard score (Z-score) can be employed to find outliers. However, it is difficult to account for the impact of sample size unless the sample results are weighted according to their size, but there is no unified form of weighting, and thresholds must be set. Methods that require setting thresholds lack persuasiveness because conclusions may differ with different thresholds.

The method proposed in this paper uses the probability density function to consider the impact of sample size and defines outliers for multiple random sampling results without setting thresholds. For a set of probability density functions corresponding to multiple sampling results, under the definition of this work, there may be no outliers, one outlier or multiple outliers. Sometimes all probability density functions in the set are outliers, resembling a ``fragmented'' set, which indicates extremely unstable sampling quality and cannot give reliable results.

\section{Method for finding outliers}

When there is no selection effect, assuming N events are detected, the more samples (n), the more reliable $\theta=N/n$. However, $\theta$ cannot reflect n, so a probability density function is needed to replace $\theta$. The larger n, the smaller the information entropy\cite{6773024} of the corresponding probability density function. Now suppose we want to investigate how many stars in a sky area are giants; we perform spectroscopic observations of that region and obtain spectra for n stars, analyzing and finding that N of them are giants. The probability of finding a giant in this sky area fits the binomial distribution

\begin{equation}
p(N \mid \theta,n)= \theta^{N}(1-\theta)^{n-N}
\label{b}
\end{equation}

Equation \ref{b} indicates that even if the proportion of giants ($\theta$) in this sky area is constant, the probability of finding $n\theta$ giants from $n$ stars in that sky area is not equal to 1, which is consistent with the Theorem of Large Numbers\cite{bernoulli1713jacobi,khintchine1936legge,loeve1977elementary}. In Bayesian Statistics, we have

\begin{equation}
p(\theta \mid N,n) \propto p(N \mid \theta,n) p(\theta)
\end{equation}

A prior distribution $p(\theta)$ is required, but we know nothing about it. For example, in this scenario, we need to test the galactic model using the proportion of giants, so we cannot correct the observational results based on the prior parameters obtained from the model. Besides, during astronomical observations, we are bound to see more giants because they are brighter than non-giants (turn-off stars) at the same distance. Therefore, we do not expect to obtain a truly complete $\theta$; it is good enough to be complete within a certain brightness (magnitude) range. However, sometimes observers also tend to select stars that are either bluer or redder, and the color distribution of giants differs from that of non-giants. Thus, color bias can affect the proportion of giants. Even if the observer's color bias is known, it is difficult to quantify its impact on the proportion of giants. In summary, the prior distribution cannot be estimated, so the prior distribution is assumed to be an uniform distribution

\begin{equation}
p(\theta)=\left\{\begin{array}{ll}
1 & 0<\theta<1 \\
0 & \text { otherwise }
\end{array}\right.
\end{equation}

then we have

\begin{equation}
p(\theta \mid N,n) \propto \theta^{N}(1-\theta)^{n-N}
\end{equation}

To ensure $\int_{0}^{1} p(\theta \mid N) d \theta=1$, it has been proved that

\begin{equation}
p(\theta \mid N,n) = \text{Beta}(\theta \mid N+1, n-N+1)
\label{p}
\end{equation}

where

\begin{equation}
\text{Beta}(\theta \mid a,b)=\frac{\Gamma (a+b)\theta^{a-1}(1-\theta)^{b-1}}{\Gamma (a)\Gamma(b)} 
\end{equation}

where $a=N+1$, $b=n-N+1$, and

\begin{equation}
\Gamma(x)=\int_{0}^{+\infty} t^{x-1} e^{-t} \mathrm{~d} t
\end{equation}

If there is only one sampling result, it ends here. However, in practical situations, the conditions during sampling are always changing. Even if it is unclear whether these specific conditions will actually lead to selection effects, the sampling result should be divided into multiple sampling results based on these conditions. In this scenario, there are always multiple observations of the same sky area, and in each observation the selection biases differ. For instance, one observation might be biased towards bluer stars, another towards redder stars, and another might even have undergone pre-filtering to exclude giants, albeit with a pre-filter accuracy that is not one hundred percent, leaving a small number of giants behind. As a result, even if an observation provides a vast sample size, it can still be unreliable, whereas a result with a much smaller sample size might actually be closer to the truth. If sampling results with strong selection biases are not treated as outliers and removed, it will inevitably lead to biases in the overall probability density function. Here comes the definition of outliers without setting thresholds.

Now assuming this sky area has been observed k times (k>3), then we have $N_1$, $N_2$, $N_3$... $N_k$ and $n_1$, $n_2$, $n_3$... $n_k$. So,

\begin{equation}
p_i(\theta)=p(\theta \mid N_i,n_i)=\text{Beta}(\theta \mid N_i+1,n_i-N_i+1)
\end{equation}

If, in a few sampling results, a non-uniform prior distribution assumption is used, resulting in a different form of the corresponding $p_i(\theta)$ from that in Equation \ref{p}, this is acceptable and will not affect the following definitions.

Now we have $\text{Obs}_k=\{ p_1, p_2, ..., p_k \}$. Make sure no repeated elements in the set $\text{Obs}_k$. Then define Similarity $S$,

\begin{equation}
S_i^j=\int_{0}^{1} \text{min}(p_i(\theta),p_j(\theta)) d \theta
\end{equation}

Then define 
\begin{equation}
    S_{list}(\text{Obs}_k)=\{ S_1^2, S_1^3, ... S_1^k, S_2^3, S_2^4, ... S_2^k, ... S_{k-1}^{k} \} 
\end{equation}

where $|S_{list}(\text{Obs}_k)|=1/2 \cdot k \cdot (k-1)$. Define

\begin{equation}
    \text{min}_1(\{ A \})=\text{min}(\{ A \})
\end{equation}

and

\begin{equation}
    \text{min}_2(\{ A \})=\text{min}(\{ A \} \verb|\| \text{min}_1(\{ A \}))
\end{equation}

so we have

\begin{eqnarray}  
    &&\text{min}_x(\{ A \}) = \text{min}(\{ A \} \verb|\| \\  
    &&\{ \text{min}_1(\{ A \}), \text{min}_2(\{ A \}), ..., \text{min}_{x-1}(\{ A \}) \}  \nonumber)  
\end{eqnarray}

Then define 

\begin{eqnarray}
&&\text{MIN}_{n}(\{ A \}) \\
&& =\{ \text{min}_1(\{ A \}), \text{min}_2(\{ A \}), ... \text{min}_{n}(\{ A \}) \} \nonumber \\
&&(n<|\{ A \}|) \nonumber
\end{eqnarray}

and

\begin{eqnarray}
    &&Checklist(S_{list}(\text{Obs}_k)) \\
    &&= \text{MIN}_{k-1}(S_{list}(\text{Obs}_k)) \nonumber \\
    &&=\{S_{a_1}^{b_1},S_{a_2}^{b_2}, ... S_{a_{k-1}}^{b_{k-1}} \}  \nonumber
\end{eqnarray}

and a check function

\begin{equation}
Check(x \mid S_a^b)=\left\{\begin{array}{ll}
1 & \quad a=x | b=x \\
0 & \quad \text { otherwise }
\end{array}\right.
\end{equation}

and an operator

\begin{equation}
    \hat{C}_i(\{ A \})= \{ Check(i \mid a) \mid a \in A) \}
\end{equation}
then define

\begin{equation}
    Unsi(i \mid \text{Obs}_k)=\sum \hat{C}_i(Checklist(S_{list}(\text{Obs}_k)))
\end{equation}

if observation i is an outlier in k observations, we have

\begin{equation}
    Unsi(i \mid \text{Obs}_k)=|Checklist(S_{list}(\text{Obs}_k))|=k-1
\end{equation}

we can also define an operator

\begin{eqnarray}
    && \\
    && \hat{Out}_1(\text{Obs}_k) \nonumber \\
    =&& \left\{\begin{array}{ll} p_i & \ Unsi(i \mid \text{Obs}_k)=|\text{Obs}_k|-1 \\
None & \ Unsi(a \mid \text{Obs}_k)<|\text{Obs}_k|-1, \ \forall 1 \leq a \leq k
\end{array}\right. \nonumber
\end{eqnarray}

and

\begin{equation}
    \hat{Out}_2(\text{Obs}_k)=\hat{Outlier}_1(\text{Obs}_k \verb|\| \hat{Out}_1(\text{Obs}_k))
\end{equation}

then we have

\begin{eqnarray}
    && \\
    && \hat{Out}_n(\text{Obs}_k) \nonumber \\
    &&=\hat{Out}_1(\text{Obs}_k \verb|\| \nonumber \\
    &&\{ \hat{Out}_1(\text{Obs}_k)), \hat{Out}_2(\text{Obs}_k)), ..., \hat{Out}_{n-1}(\text{Obs}_k) \}) \nonumber
\end{eqnarray}

If $\hat{Out}_n$ is undefined, $\hat{Out}_m, \ \forall m > n$ are undefined. If $\hat{Out}_{k-3}(\text{Obs}_k)$ is defined, $\hat{Out}_{k}(\text{Obs}_k)$ is defined so $\text{Obs}_k$ is ``fragmented'', no reliable results should be given by a ``fragmented'' set.

\begin{figure}[h]
\includegraphics[width=0.45\linewidth]{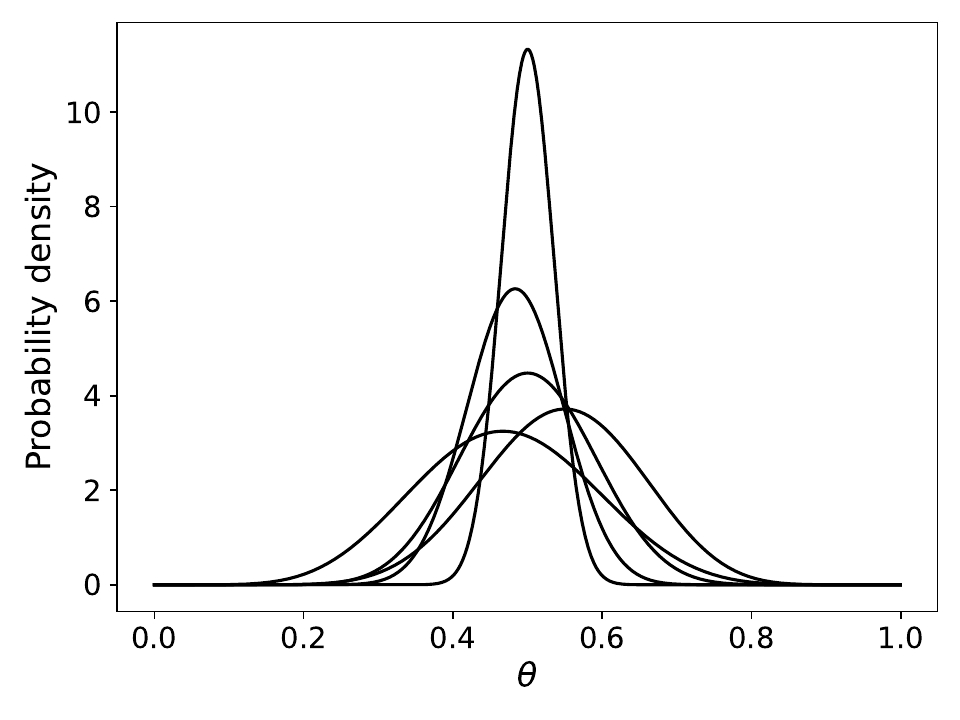}
\includegraphics[width=0.45\linewidth]{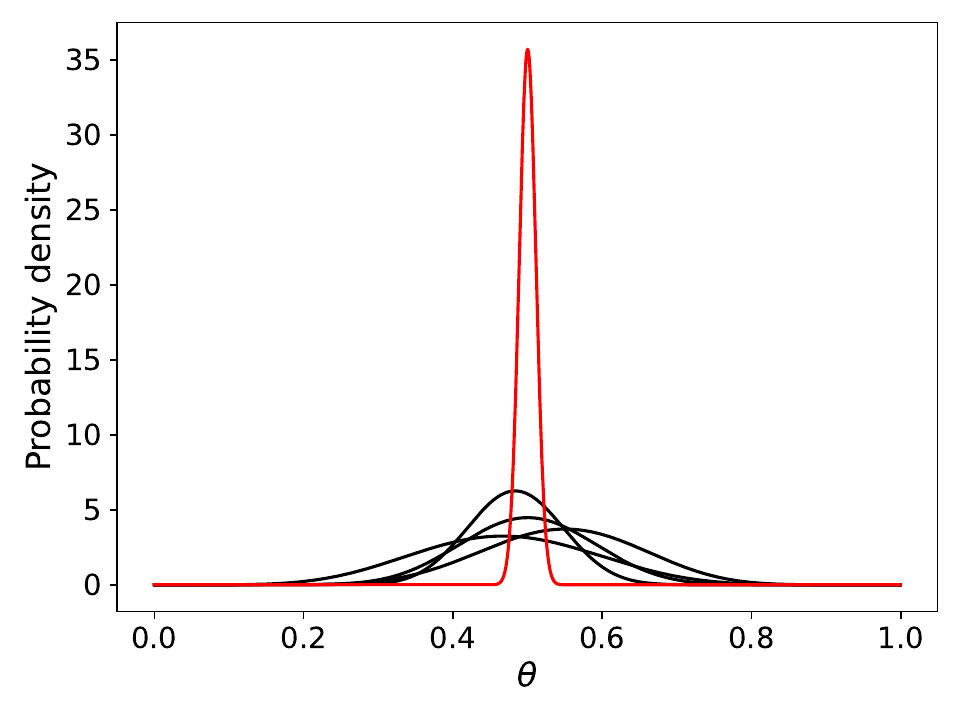}
\includegraphics[width=0.45\linewidth]{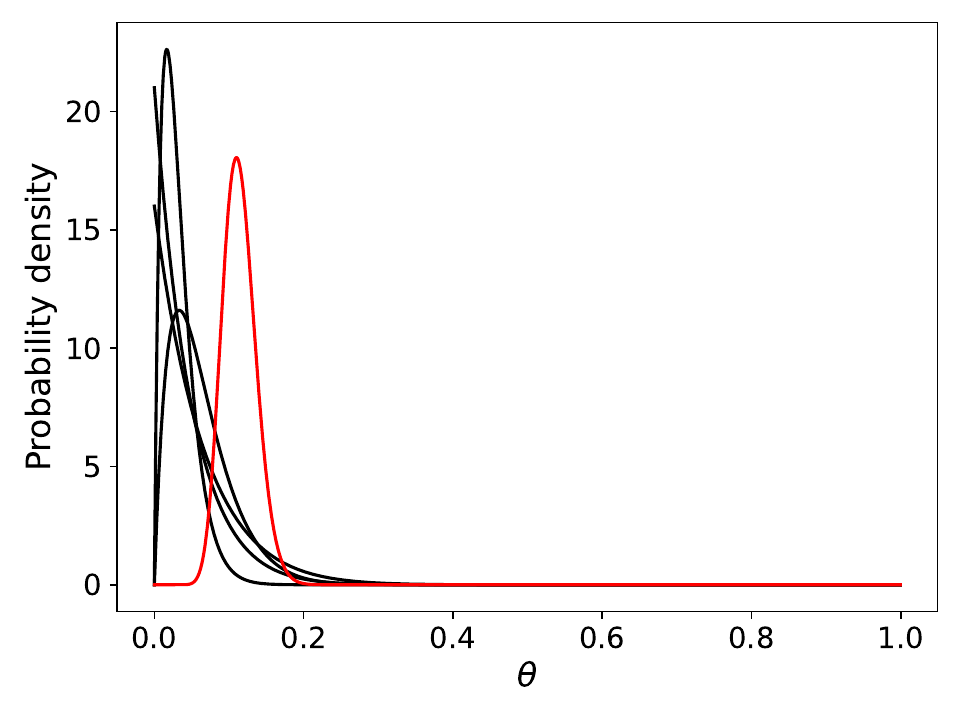}
\includegraphics[width=0.45\linewidth]{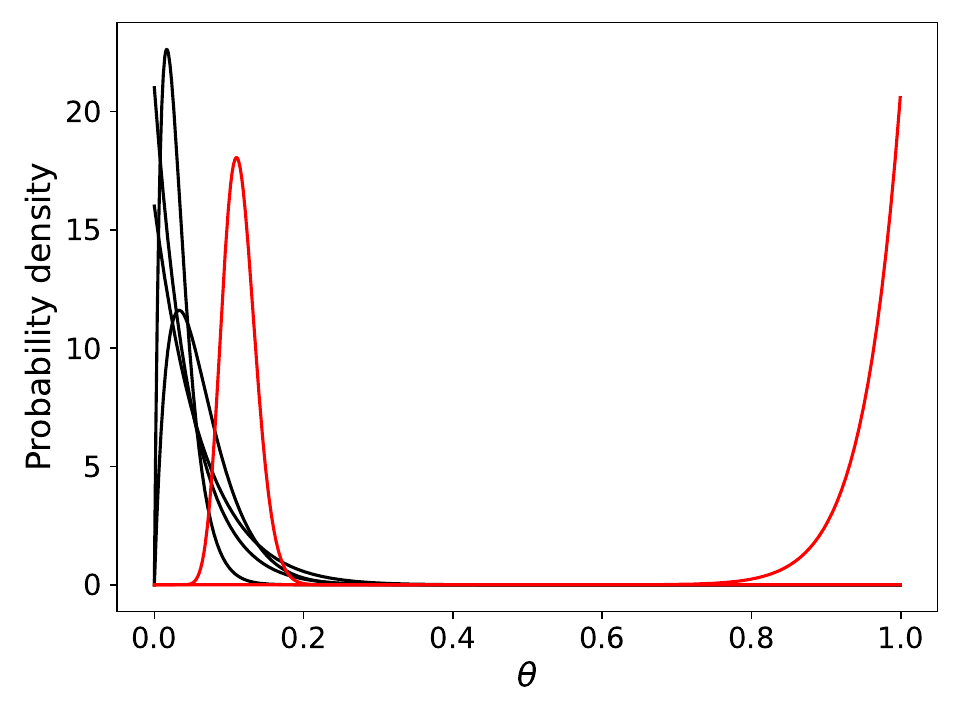}
\caption{All curves in the four panels are probability density distributions given by synthetic observations. Each observation corresponds to a curve. Red curves are outliers. From left to right and from top to bottom, the first panel shows an example of five observations without outliers, the second panel shows a very sharp outlier compared with other curves, the third panel shows a common type of outlier, and the fourth panel shows the case with two outliers. The pictures can be drawn by the code provided in the appendix.}
\label{example}
\end{figure}

Figure~\ref{example} are some examples.

\begin{supplement}
\stitle{Python code}
\sdescription{The code for drawing the sketch map is also included}

\begin{verbatim}
import numpy as np
from scipy.stats import beta
from collections import Counter
def S(Ni,ni,Nj,nj):
    h=0.001#integration step size
    theta=np.arange(0,1,h)
    pi=beta.pdf(theta,Ni+1,ni-Ni+1)
    pj=beta.pdf(theta,Nj+1,nj-Nj+1)
    minij=[min(a,b) for a,b in zip(pi,pj)]
    minij=np.array(minij)
    return np.sum(minij*h)
def out(N,n):
    k=len(N)
    Slist=[]
    ilist=[]
    jlist=[]
    for i in range(k):
        for j in range(k):
            if i>=j:
                continue
            else:
                Slist.append(
                S(N[i],n[i],N[j],n[j])
                )
                ilist.append(i)
                jlist.append(j)
    ilist=np.array(ilist)
    jlist=np.array(jlist)
    Slist=np.array(Slist)
    ilist=ilist[np.argsort(Slist)]
    jlist=jlist[np.argsort(Slist)]
    ilist=ilist[:k-1].tolist()
    jlist=jlist[:k-1].tolist()
    l=ilist+jlist
    counter = Counter(l)
    mce, mcc = counter.most_common(1)[0]
    if mcc<k-1:
        return -1
    else:
        return mce
    
def main(N,n):
    if len(N)!=len(n):
        print('len(N)!=len(n)')
        return -1
    d=np.array(n)-np.array(N)
    if len(d[d<0])>0:
        print("n < N")
        return -1
    outN=[]
    outn=[]
    output=len(N)+1
    while output>=0:
        output=out(N,n)
        if output>=0:
            outliersN.append(N[output])
            outliersn.append(n[output])
            N=N[:output]+N[output+1:]
            n=n[:output]+n[output+1:]
            if len(N)==1:
                print('Fragmented!')
                outN.append(N[0])
                outn.append(n[0])
                return N,n,outN,outn
        else:
            return N,n,outN,outn



#example:
import matplotlib.pyplot as plt
N=[15,11,7,29,100]
n=[30,20,15,60,200]
newN,newn,outN,outn=main(N,n)  
print(outN)
print(outn)
theta=np.arange(0,1,0.001)    
for i in range(len(newN)):
    plt.plot(theta, \
    beta.pdf(theta,newN[i]+1, \
    newn[i]-newN[i]+1),color='black')
for i in range(len(outN)):
    plt.plot(theta, \
    beta.pdf(theta,outN[i]+1, \
    outn[i]-outN[i]+1),color='red')
plt.xlabel('$\\theta$',fontsize=16)
plt.ylabel('Probability density', \
    fontsize=16)
plt.xticks(fontsize=14)
plt.yticks(fontsize=14)
plt.tight_layout()
plt.show()
    
\end{verbatim}

\end{supplement}

\bibliographystyle{imsart-number} 
\bibliography{ref}       

\begin{thebibliography}{5}

\bibitem{2008Bayes}
\begin{barticle}[author]
\bauthor{\bsnm{Annis},~\bfnm{David~H.}\binits{D.~H.}}
(\byear{2008}).
\btitle{Bayes Linear Statistics: Theory and Methods}.
\bjournal{Journal of the American statistical association}
\bvolume{103}
\bpages{p.1319}.
\end{barticle}
\endbibitem

\bibitem{bernoulli1713jacobi}
\begin{bbook}[author]
\bauthor{\bsnm{Bernoulli},~\bfnm{Jakob}\binits{J.}}
(\byear{1713}).
\btitle{Jacobi Bernoulli,... Ars conjectandi, opus posthumum. Accedit Tractatus de seriebus infinitis, et epistola Gallice scripta De ludo pilae reticularis}.
\bpublisher{impensis Thurnisiorum, fratrum}.
\end{bbook}
\endbibitem

\bibitem{khintchine1936legge}
\begin{barticle}[author]
\bauthor{\bsnm{Khintchine},~\bfnm{A~Ya}\binits{A.~Y.}}
(\byear{1936}).
\btitle{Su una legge dei grandi numeri generalizzata}.
\bjournal{Giorn. Ist. Ital. Attuari}
\bvolume{7}
\bpages{365--377}.
\end{barticle}
\endbibitem

\bibitem{loeve1977elementary}
\begin{bbook}[author]
\bauthor{\bsnm{Lo{\`e}ve},~\bfnm{Michel}\binits{M.}} \AND \bauthor{\bsnm{Lo{\`e}ve},~\bfnm{M}\binits{M.}}
(\byear{1977}).
\btitle{Elementary probability theory}.
\bpublisher{Springer}.
\end{bbook}
\endbibitem

\bibitem{6773024}
\begin{barticle}[author]
\bauthor{\bsnm{Shannon},~\bfnm{C.~E.}\binits{C.~E.}}
(\byear{1948}).
\btitle{A mathematical theory of communication}.
\bjournal{The Bell System Technical Journal}
\bvolume{27}
\bpages{379-423}.
\bdoi{10.1002/j.1538-7305.1948.tb01338.x}
\end{barticle}
\endbibitem

\end{thebibliography}

\end{document}